\documentclass[twocolumn,showpacs,preprintnumbers,amsmath,amssymb]{revtex4}

\usepackage{graphicx}
\usepackage{bm}

\begin{document}

\title{Stabilization of decaying states and quantum vortices}

\author{K. Krajewska and J. Z. Kami\'nski}

\affiliation{Institute of Theoretical Physics, Warsaw University,
Ho\.za 69, 00-681 Warszawa, Poland}


\begin{abstract}
A two-dimensional model of an electron moving under the influence
of an attractive zero-range potential as well as external magnetic
and electric fields is analyzed. We prove by numerical
investigations that there are formed such resonances which
manifest a peculiar dependance on the electric field intensity,
i.e., although the electric field increases, the lifetime of these
resonance states grows up. It is explained that this phenomenon,
called further \textit{the stabilization}, is a close consequence
of \textit{quantum-mechanical vortices} induced by the magnetic
field and controlled by the electric field strength. In order to
get more information about these vortices the phase of
wavefunctions as well as the probability current for these stable
resonances are investigated.
\end{abstract}

\maketitle


\section{Introduction}
\label{sec:Introduction}

Applying a classical model for two-dimensional motion of electrons
in crossed magnetic and electric fields interacting, in addition,
with an repulsive short-range potential, Berglund et. al. have
shown in \cite{BHHP96} that there are such electron trajectories
which remain bound in space, i.e., which describe classical bound
states of electrons in such a configuration of electromagnetic
fields. From the quantum mechanical viewpoint this cannot happen
since a non-vanishing electric field transforms quantum bound
states into resonances with a finite lifetime. Such a problem has
been further discussed in \cite{HL99,GM99,KK1}. In particular,
Hauge and van Leeuwen \cite{HL99} have considered the case of a
repulsive `\textit{point interaction}' showing the existence of
long-living resonance states. It appears, however, that it is not
possible to define such an interaction without adopting any
additional conjectures. In order to avoid this difficulty and to
analyze a model which is mathematically well-defined, Gyger and
Martin \cite{GM99} have studied an attractive point interaction
and shown that for weak electric fields there exist long-living
resonance states, the lifetime of which, however, monotonically
decrease with an increasing electric field strength. In our recent
paper \cite{KK1} we have considered the case of an arbitrary
electric field for an attractive point interaction. We have shown
there that for a vanishing electric field there appear bound
states, the positions of which depend on the magnetic field
intensity and strength of a point interaction, measured by the
energy, $E_B$, of the only bound state supported by this
interaction. Moreover, it has turned out that these bound states
are located between the Landau levels and that there is only one
such a state per one Landau level. Our findings are in full
agreement with the results presented in \cite{CC98}. However, it
appears that for an increasing electric field the lifetime of such
a state continuously decreases which means it cannot be considered
as a quantum analogue of a classical bound state discussed in
\cite{BHHP96}, at least for non-perturbative electric fields. We
have also proven \cite{KK1} that the electric field generates new
resonance states that emerge from Landau levels, the number of
which is equal to the principal quantum number $n$, with $n$
labelling energies of Landau levels. By analyzing these states we
have managed to show that for some particular electric field
intensities and strengths of the point interaction there exist
very long-living resonances which, in our opinion, correspond to
the classical situation discussed above, i.e., the appearance of
which substantially depends on both the electric field intensity
and bound state energy $E_B$. The purpose of our paper is to
investigate these particular states and to show that their
existence is closely related to quantum vortices. In order to make
this we shall apply the notation and results already presented
in \cite{KK1}. In particular, we shall use new units in which the
energy, electric field intensity and length are scaled with
$2/\omega$, $|e|/\sqrt{m^*{\omega}^3}$ and $\sqrt{m^*\omega}$,
where $\omega=|e|\mathcal{B}/m^*$ and $m^*$ denote the cyclotron
frequency and electron's effective mass, respectively.
Throughout this paper we put $\hbar=1$ and indicate the scaled
quantities by the '\textit{tilda}' symbol.

\section{Stabilization of resonances}
\label{sec:stabilization}

Our point of departure is the exact Green's function derived in
\cite{KK1}. In our scaled units this function adopts the form
\begin{equation}
\tilde{G}( \tilde{ \mathbf{r}},\tilde{ \mathbf{r}}';\tilde{E})=
\tilde{G}_0(\tilde{ \mathbf{r}},\tilde{ \mathbf{r}}';\tilde{E})+
\frac{\tilde{G}_0(\tilde{ \mathbf{r}},\mathbf{0};\tilde{E})
\tilde{G}_0(\mathbf{0},\tilde{
\mathbf{r}}';\tilde{E})}{\tilde{D}(\tilde{ E})} \, . \label{e2.1}
\end{equation}
where
\begin{eqnarray}
\tilde{G}_0(\tilde{\mathbf{r}},\tilde{\mathbf{r}}';\tilde{E}) =
-\int_0^{\infty} \mathrm{d}s
\frac{\mathrm{e}^{\mathrm{i}(\tilde{E}+\mathrm{i}\varepsilon)s}}
{\sin s} \nonumber\\
\times\exp\Biggl[ \frac{\mathrm{i}}{4}\Bigl(
(\tilde{x}-\tilde{x}')^2 + (\tilde{y}-\tilde{y}')^2\Bigr )\cot s
+ \frac{\mathrm{i}}{2}(\tilde{x}\tilde{y}'-\tilde{x}'\tilde{y})
\nonumber\\
-\mathrm{i}\tilde{\mathcal{E}}s(\tilde{x}+\tilde{x}') +
\mathrm{i}\tilde{\mathcal{E}}\bigl( s \cot s-1\bigr ) \bigl(
\tilde{y}-\tilde{y}'+\tilde{\mathcal{E}} s\bigr )\Biggr ] \, ,
\label{e2.2}
\end{eqnarray}
and
\begin{eqnarray}
\tilde{D}(\tilde{E})=\ln\biggl(
\frac{\tilde{E}_B}{\tilde{E}}\biggr )+ \int_0^{\infty}\mathrm{d}s
\mathrm{e}^{\mathrm{i}(\tilde{E}+ \mathrm{i}\varepsilon)s} \nonumber\\
\biggl(\frac{\exp\bigl( \mathrm{i} {\tilde\mathcal{E}}^2 s(s\cot
s-1)\bigr )}{\sin s}-\frac{1}{s}\biggr ) \, . \label{e2.3}
\end{eqnarray}
\begin{figure}
\begin{center}
\includegraphics{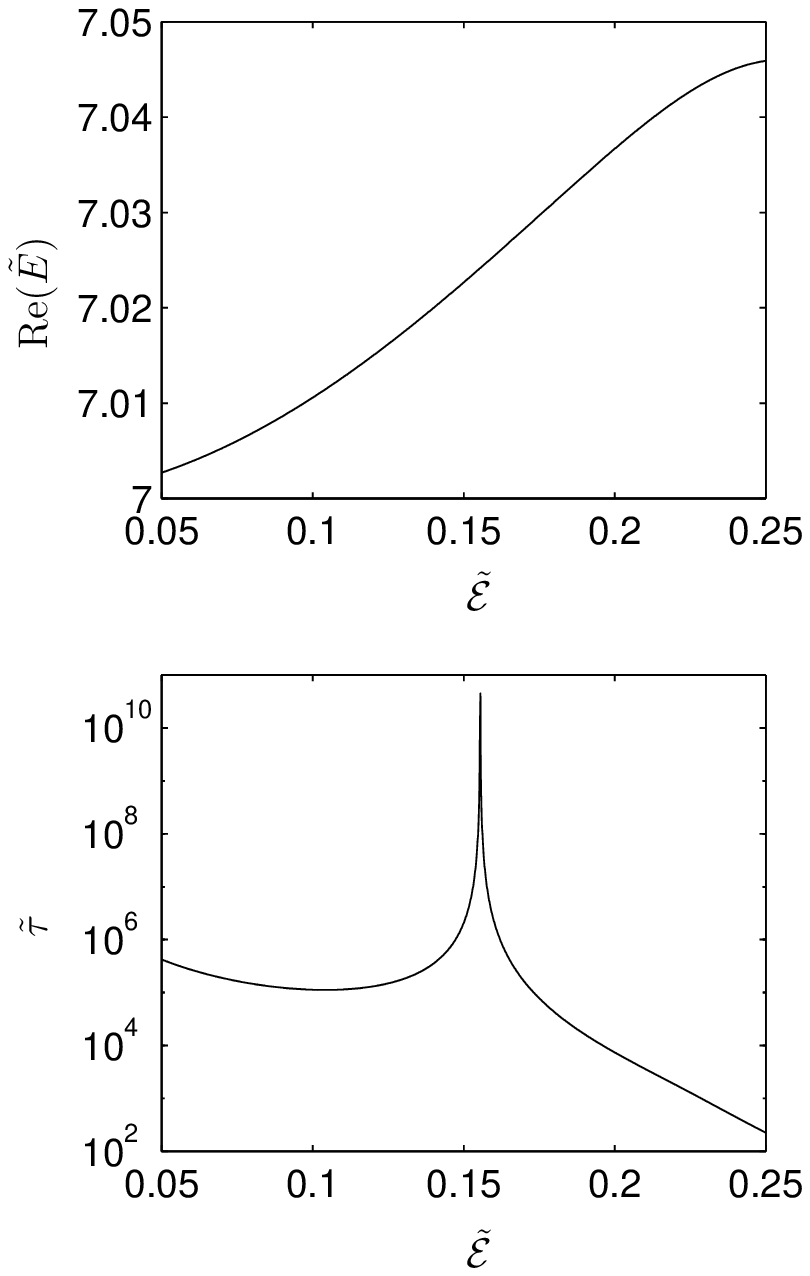}
\end{center}
\caption{Stabilization effect for one of the
electric-field-induced impurity states.} \label{f1}
\end{figure}
\begin{figure}
\begin{center}
\includegraphics{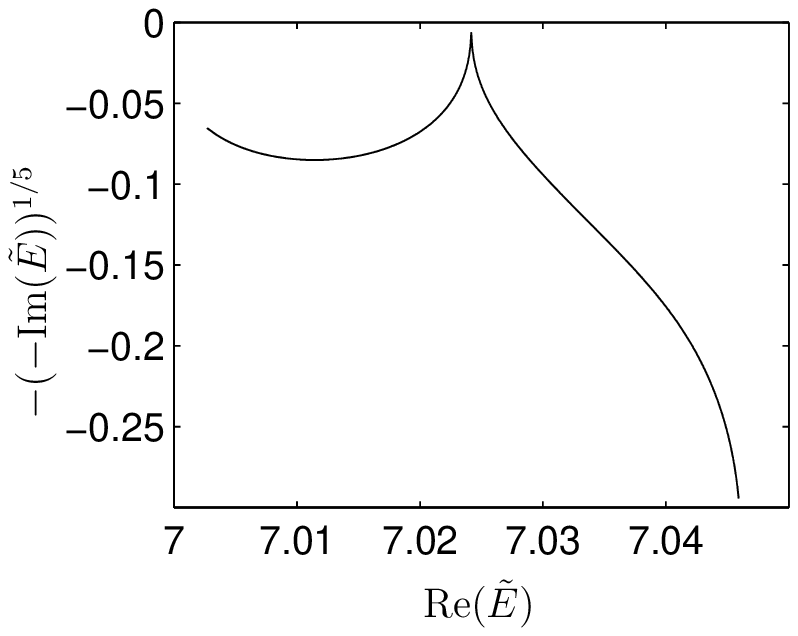}
\end{center}
\caption{Path of the Green's function pole with a changing
electric field.} \label{f2}
\end{figure}
As has been expected, for the vanishing point interaction, i.e.
when $\tilde{E}_B=0$, the exact Green's function reduces to the
Green's function for electrons moving only in crossed magnetic and
electric fields. Zeros of $\tilde{D}(\tilde{E})$ determine complex
energies of those resonances which are induced by the presence of
an impurity described by an attractive point interaction. On top
of that, the exact Green's function allows us to establish
unambiguously the exact form of the wavefunctions for such a
resonance provided that it is not degenerate. Indeed, if
$E_\mathrm{r}$ is the resonance energy, hence the wavefunctions
are defined by the relation (not in our scaled units)
\begin{equation}
\lim_{E\rightarrow E_\mathrm{r}}(E-E_\mathrm{r})
G(\mathbf{r},\mathbf{r}';E)=
\psi_\mathrm{r}(\mathbf{r})[\psi^{A}_\mathrm{r}(\mathbf{r}')]^* \,
. \label{e2.4}
\end{equation}
At this point it is crucial to stress that for resonances, in
contrast to the bound state problem, there exist two wavefunctions
which we call the retarded $\psi_\mathrm{r}(\mathbf{r})$ and the
advanced $\psi^{A}_\mathrm{r}(\mathbf{r})$ wavefunctions (see also
\cite{F87,Reson,MFSF00}). For these functions the normalization
condition has the form
\begin{equation}
\int \, \mathrm{d}\mathbf{r} \, \psi_\mathrm{r}(\mathbf{r})
[\psi^{A}_\mathrm{r}(\mathbf{r})]^* = 1 \, . \label{e2.5}
\end{equation}
It turns out that for an infinitely long-living resonance, i.e.
when the energy becomes real, both these wavefunctions are
identical. This property is exemplified by our model. Defining, in
our scaled units, the un-normalized retarded and advanced
wavefunctions as
\begin{equation}
\tilde{\psi}_\mathrm{r}(\tilde{\mathbf{r}})=-\mathrm{i} \,
\tilde{G}_0(\tilde{\mathbf{r}},\mathbf{0};\tilde{E}_\mathrm{r}) \,
, \label{e2.6}
\end{equation}
and
\begin{equation}
[\tilde{\psi}^{A}_\mathrm{r}(\tilde{\mathbf{r}})]^* =\mathrm{i} \,
\tilde{G}_0(\mathbf{0},\tilde{\mathbf{r}};\tilde{E}_\mathrm{r}) \,
, \label{e2.7}
\end{equation}
one can prove that
\begin{equation}
\int \, \mathrm{d}\tilde{\mathbf{r}} \,
\tilde{\psi}_\mathrm{r}(\tilde{\mathbf{r}})
[\tilde{\psi}^{A}_\mathrm{r}(\tilde{\mathbf{r}})]^* =
\frac{\partial \tilde{D}(\tilde{E})}{\partial
\tilde{E}}\mid_{\tilde{E}=\tilde{E}_\mathrm{r}}
=\tilde{D}'(\tilde{E}_\mathrm{r}) \, . \label{e2.8}
\end{equation}
Hence, in the close vicinity of a resonance,
where $\tilde{D}(\tilde{E}_\mathrm{r})=0$, we get the following
relation
\begin{equation}
\tilde{G}(\tilde{\mathbf{r}},
\tilde{\mathbf{r}}';\tilde{E})\cong\frac{1}{\tilde{E}-\tilde{E}_r}
\frac{\tilde{\psi}_\mathrm{r}(\tilde{\mathbf{r}})
[\tilde{\psi}^{A}_\mathrm{r}(\tilde{\mathbf{r}}')]^*}
{\tilde{D}'(\tilde{E}_\mathrm{r})} \, , \label{e2.9}
\end{equation}
which agrees with (\ref{e2.4}) and (\ref{e2.5}). Furthermore, for
real $\tilde{E}_\mathrm{r}$ both
$\tilde{\psi}_\mathrm{r}(\tilde{\mathbf{r}})$ and
$\tilde{\psi}^{A}_\mathrm{r}(\tilde{\mathbf{r}})$ are identical,
which we have also checked numerically.
\begin{figure}
\begin{center}
\includegraphics{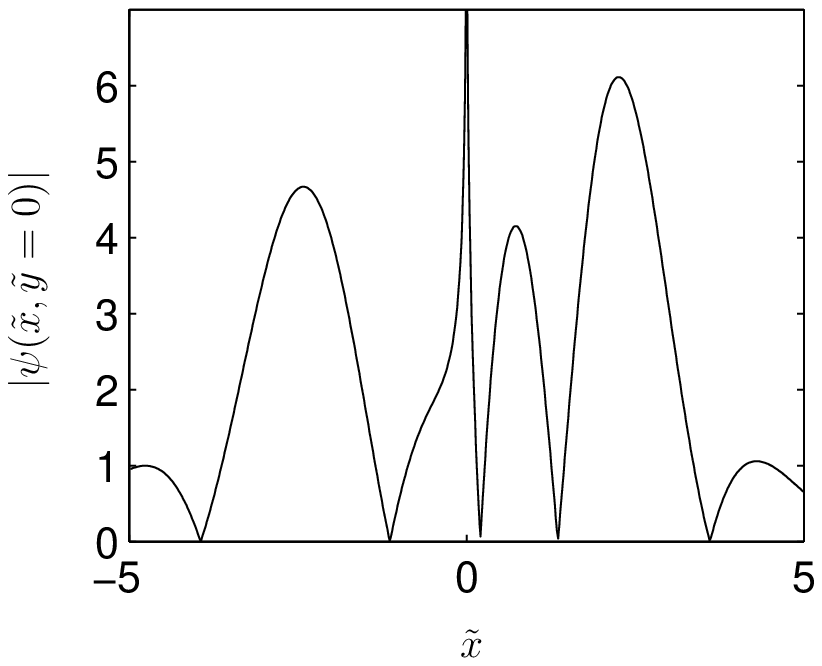}
\end{center}
\caption{Modulus of $\psi(\tilde{x},\tilde{y}=0)$ for the stable
resonance.} \label{f3}
\end{figure}

We have already pointed out in \cite{KK1} that the
electric-field-induced resonances exhibit peculiar behavior with
an increasing electric field strength. Namely, for some particular
values of electric fields and binding energies the lifetime of
these states (within a computational error) can be infinite, i.e.,
we observe the stabilization process. We have checked that this
phenomenon cannot happen for the 'old' localized impurity-induced
states already discussed in \cite{GM99,CC98}. For this reason let
us consider the electric-field-induced resonances located near the
third excited Landau level of the scaled energy $\tilde{E}=7$.
There are three such states and the stabilization phenomenon for
one of them is presented in Fig. \ref{f1}. It happens for the
scaled electric field $\tilde{\mathcal{E}}=0.1555$ and binding
energy $\tilde{E}_B=-6.4$ (for other values of these parameters a
very similar behavior occurs for the remaining two resonances,
therefore, we shall not attach the corresponding figures here). In
Fig. \ref{f2} we present positions of energies in the complex
energy plane for this resonance with changing electric field. For
the visual purpose we have raised the imaginary part of this
energy to the power $1/5$. We observe that initially an increasing
electric field tears off the resonance energy from the Landau
level (as it has been demonstrated in \cite{KK1}) and pushes it
down. However, with a still increasing electric field strength the
energy starts migrating back towards the real axis and approaches
it, within a numerical error, for some particular value of
$\tilde{\mathcal{E}}$, for which the lifetime of this resonance
becomes infinite. For still larger electric fields the energy
again, as expected, withdraws downwards from the real axis. What
is the reason for this peculiar behavior? In order to answer this
question we have analyzed the wavefunction of this resonance. We
present in Fig. \ref{f3} the wavefunction for $\tilde{y}=0$ when
the stabilization takes place. It is clearly seen that at the
origin $\tilde{\mathbf{r}}=\mathbf{0}$, where the impurity is
located, the wavefunction is singular and one can prove by
analyzing equation (\ref{e2.6}) that this singularity is of the
logarithmic type. However, more interesting is that the
wavefunction posses five zeros which are attributed to
quantum-mechanical vortices, as it will follow below. This means
that responsible for the stabilization effect is the particular
vortex-type motion of the `probability fluid' and, hence, induced
by such a motion quantum-mechanical interferences.
\begin{figure}
\begin{center}
\includegraphics[width=8.5cm]{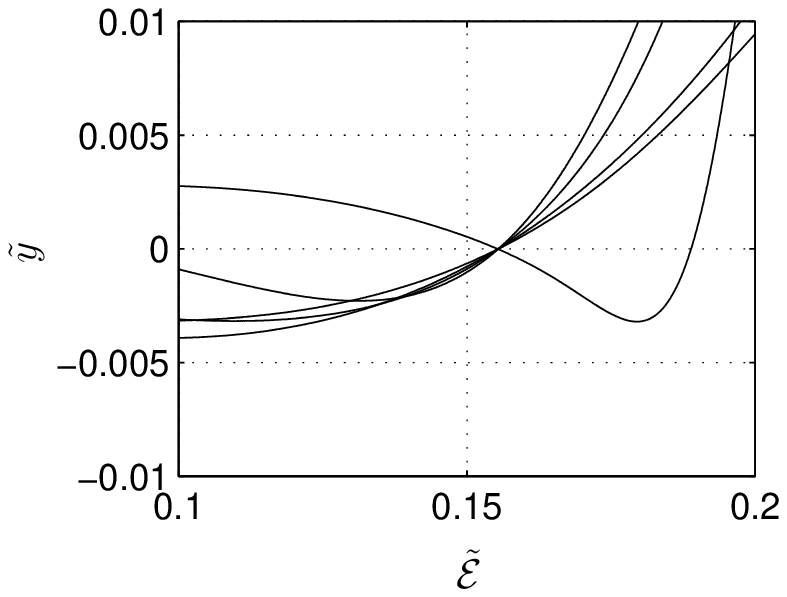}
\end{center}
\caption{$\tilde{y}$-coordinates of vortices as functions of the
electric field strength.} \label{f4}
\end{figure}

\section{Vortices}
\label{sec:vortices}

\begin{figure}
\begin{center}
\includegraphics{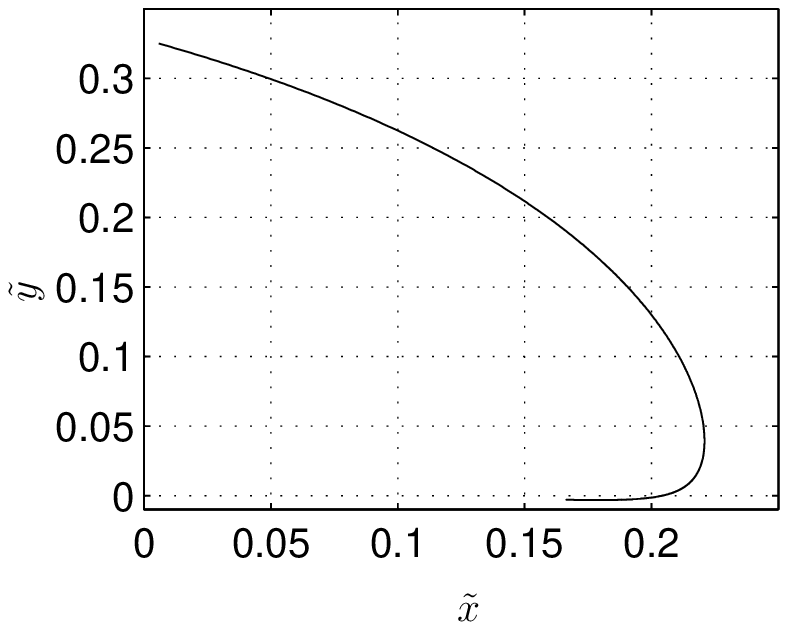}
\end{center}
\caption{Path in the coordinate plane of a vortex for different
values of $\tilde{\mathcal{E}}$.} \label{f5}
\end{figure}

It is well-known that the magnetic field induces
quantum-mechanical vortices and that the circulation around them
is quantized \cite{BCK92,BBS00}. The same phenomenon occurs in our
model, except the fact that instead of the vortex line for
three-dimensional space we have the vortex point for two
dimensions. Moreover, it turns out that the electric field can be
used as a control parameter with the help of which one can change
the positions of these vortices. In particular, it might happen
that for some values of the electric field strength all these
vortices can be aligned along the axis defined by the direction of
the electric field vector, in our case along the $x$-axis. If this
takes place the stabilization occurs, i.e., the electric field is
not able to push out the electron from the vicinity of an
impurity. In order to show this we have determined zeros of the
wavefunction $\tilde{\psi}(\tilde{\mathbf{r}})$ for
$\tilde{E}_B=-6.4$ and for different electric field strengths. The
$\tilde{y}$-coordinates of five zeros for the resonance state
considered above are presented in Fig. \ref{f4}. We see that they
cross the $\tilde{x}$-axis exactly for such an electric field for
which the stabilization appears. Moreover, in Table \ref{tabel1}
we present the $\tilde{x}$-coordinates of these vortices for this
particular electric field. The path in the coordinate plane of the
vortex number 3 from Table \ref{tabel1} is displayed in Fig.
\ref{f5}. We observe that for small electric fields this vortex is
located just below the $\tilde{x}$-axis and for a particular
electric field strength it crosses this axis (at this moment the
lifetime goes to infinity) and afterwards runs away from the
impurity center in the perpendicular direction to the electric
field vector. The very similar behavior appears for the remaining
four vortices. These numerical findings show that the alignment of
all vortices along the $\tilde{x}$-axis is the reason for the
stabilization phenomenon observed in Figs. \ref{f1} and \ref{f2}.

\begin{figure}
\begin{center}
\includegraphics{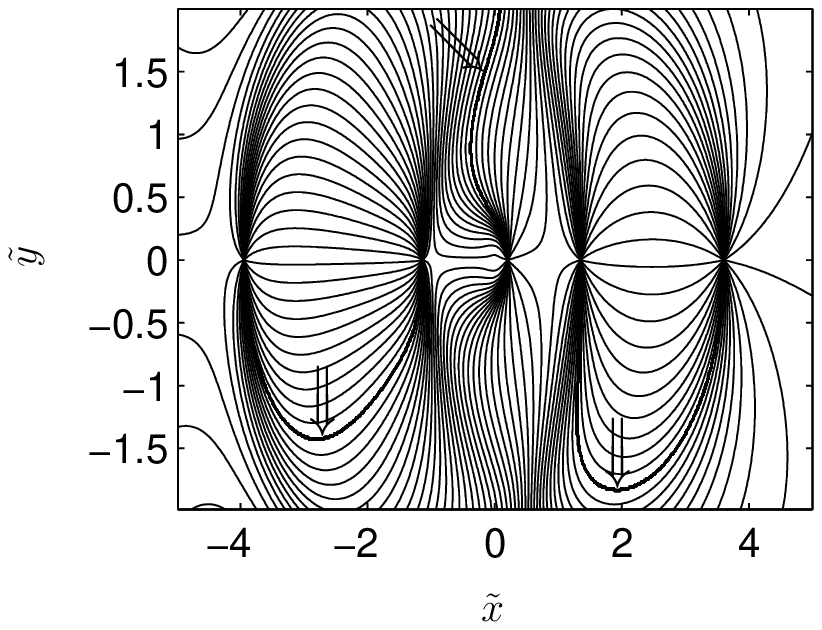}
\end{center}
\caption{Contour plot of the phase of the wavefunction for the
stable resonance.} \label{f6}
\end{figure}

\begin{figure}
\begin{center}
\includegraphics{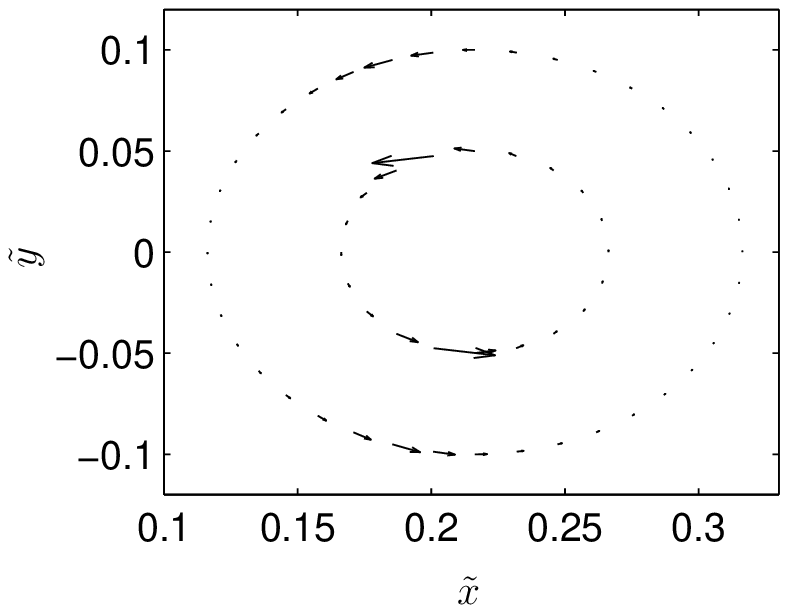}
\end{center}
\caption{Quiver plot of the probability velocity for the stable
resonance and for two chosen contours.} \label{f7}
\end{figure}

In order to get more insight into the vortex structure of stable
resonances we have to investigate the phase of the wavefunction
and probability current. Since for stable resonances the retarded
and advanced wavefunctions are identical, therefore, it is
sufficient to explore the phase of only one of them. We define the
phase $S(\tilde{\mathbf{r}})$ by the equation
\begin{equation}
\tan\bigl ( S(\tilde{\mathbf{r}})\bigr )
=\frac{\mathrm{Im}(\tilde{\psi}(\tilde{\mathbf{r}}))}
{\mathrm{Re}(\tilde{\psi}(\tilde{\mathbf{r}}))} \, , \label{e3.1}
\end{equation}
with the assumption that $-\pi < S(\tilde{\mathbf{r}})
\leqslant \pi$. The scaled probability current is equal to
\begin{equation}
\tilde{\mathbf{j}}(\tilde{\mathbf{r}})=\mathrm{Im} \bigl (
\tilde{\psi}^*(\tilde{\mathbf{r}})(\tilde{\mathbf{\nabla}} -{\rm
i} e\tilde{\mathbf{A}}(\tilde{\mathbf{r}}))
\tilde{\psi}(\tilde{\mathbf{r}})\bigr ) \, , \label{e3.2}
\end{equation}
where both $\tilde{\mathbf{\nabla}}$ and $\tilde{\mathbf{A}}$ are
scaled the nabla operator and vector potential. This current
defines the scaled probability velocity
$\tilde{\mathbf{v}}(\tilde{\mathbf{r}})$ \cite{BCK92,BBS00}
\begin{equation}
\tilde{\mathbf{j}}(\tilde{\mathbf{r}})=(\tilde{\mathbf{v}}(\tilde{\mathbf{r}})
-e\tilde{\mathbf{A}}(\tilde{\mathbf{r}}))\mid\!
\tilde{\psi}(\tilde{\mathbf{r}})\!\mid^2\, , \label{e3.3}
\end{equation}
which has to fulfil the quantization condition
\begin{equation}
\Gamma_C=\oint_C \tilde{\mathbf{v}}(\tilde{\mathbf{r}}) \cdot
\mathrm{d}\tilde{\mathbf{r}}=2\pi N \, , \label{e3.4}
\end{equation}
where $C$ is an arbitrary contour in the
$\tilde{x}\tilde{y}$-plane and $N$ is an integer number. We have
checked in our numerical investigations that within a numerical
error the foregoing quantization condition is fulfilled by any
contour in the $\tilde{x}\tilde{y}$-plane. In particular, by
choosing as a contour small circles encircling only one vortex we
have calculated circulations for five vortices appearing for this
state (see, Table \ref{tabel1}). We see that three of them have
positive circulations (we call them vortices), whereas the
remaining two negative ones (we call them anti-vortices). It is
interesting to notice that an anti-vortex appears in the pair with
a vortex. This fact is presented in Fig. \ref{f6} where the
contour plot of the phase is drawn. We clearly see in this plot
five vortices located on the $\tilde{x}$-axis with the
$\tilde{x}$-coordinates collected in Table \ref{tabel1}. In this
figure three arrows indicate three bold lines which correspond to
the discontinuity of the phase, where the phase suddenly changes
its value by $\pm 2\pi$. We see that the vortex and anti-vortex
are always connected by such a discontinuity line, whereas for the
remaining third vortex this line escapes to the infinity. The
quiver plot of the probability velocity vector
$\tilde{\mathbf{v}}(\tilde{\mathbf{r}})$ near this vortex is
presented in Fig. \ref{f7}. It is shown that the magnitude of
$\tilde{\mathbf{v}}(\tilde{\mathbf{r}})$ becomes larger in the
closer vicinity of the vortex point. Such a behavior guarantees
that the circulation $\Gamma_C$ is preserved indeed, i.e., the
quantization condition (\ref{e3.4}) holds for any closed contour
encircling the vortex.
\begin{table} 
\caption{Positions of five vortices and their circulations.}
\begin{center}
\begin{ruledtabular}
\begin{tabular}{crc}
vortex & $\tilde{x}\quad \quad$ & $\Gamma_C/2\pi$
\\\hline
 1 & $-$3.943428 & $-$1\\ 
 2 & $-$1.148087 & $+$1\\
 3 & 0.216306 & $+$1\\ 
 4 & 1.347754 & $+$1\\ 
 5 & 3.610649 & $-$1
\end{tabular}
\end{ruledtabular}
\end{center}
\label{tabel1}
\end{table}

To summarize, we have shown that for the electric-field-induced
impurity states there exist some particular values of
$\tilde{\mathcal{E}}$ and $\tilde{E}_B$ for which the
stabilization phenomenon occurs and that this effect is directly
connected with quantum-mechanical vortices which, for these
stabilized resonances, are all placed along the $\tilde{x}$-axis.
We have discussed this problem for one particular resonance, but
our findings are generally valid for all other states in the sense
that we have checked this for states located in the vicinity of
excited Landau levels of the principal quantum numbers $n=1$, 2
and for the case of $n=3$, studied in this paper. The
stabilization effect does not appear for impurity states discussed
previously in \cite{CC98,GM99}. It also appears that the number of
vortices present in these states is equal to $2n-1$.

\acknowledgments
This work has been supported in part by the Polish Committee
for Scientific Research (Grant No. KBN 2 P03B 039 19).


\begin{thebibliography}{99}
\bibitem{BHHP96}
N. Berglund, A. Hansen, E. H. Hauge, J. Piasecki, Phys. Rev. Lett.
77 (1996) 2149.
\bibitem{HL99}
E. H. Hauge, J. M. J. van Leeuwen, Physica A 268 (1999) 525.
\bibitem{GM99}
S. Gyger, P. A. Martin, J. Math. Phys. 40 (1999) 3275.
\bibitem{KK1}
K. Krajewska, J. Z. Kami\'nski, quant-ph/0205106.
\bibitem{CC98}
R. M. Cavalcanti, C. A. A. de Carvalho, J. Phys. A 31 (1998) 2391.
\bibitem{F87}
F. H. M. Faisal, Theory of Multiphoton Processes, New York, 1987.
\bibitem{Reson}
V. I. Kukulin, V. M. Krasnopol'sky, J. Hor\'a\v cek, Theory of
Resonances. Principles and Applications, Kluwer Akademic
Publishers, Prague, 1989.
\bibitem{MFSF00}
N. L. Manakov, M. V. Frolov, A. F. Starace, I. I. Fabrikant, J.
Phys. B 33 (2000) R141.
\bibitem{BCK92}
I. Bia\l ynicki-Birula, M. Cieplak, J. Kami\'nski, Theory of
Quanta, Oxford University Press, New York, 1992.
\bibitem{BBS00}
I. Bia\l ynicki-Birula, Z. Bia\l ynicka-Birula, C. \'Sliwa, Phys.
Rev. A 61 (2000) 032110.
\end{thebibliography}
\end{document}